\documentclass[12pt]{article}
\usepackage[utf8]{inputenc}
\usepackage{amsfonts}
\usepackage{enumitem}
\usepackage{graphicx}
\usepackage{cite}
\usepackage{color}
\usepackage{pst-node}
\usepackage{tikz-cd} 
\usepackage{amsmath,amssymb,amsthm}
\usepackage{mathrsfs}
\usepackage{slashed}
\usepackage[english]{babel}
\usepackage[margin=3cm]{geometry}

\def \be {\begin{equation}}
\def \ee {\end{equation}}
\def \bea {\begin{eqnarray}}
\def \eea {\end{eqnarray}}
\def \nn {\nonumber}

\def \rr {\raise.35ex\hbox{\small $\prime$}\kern-.17em{\mbox{\large $\imath$}}}

\def \dels {\partial\kern-.6em /\kern.1em}
\def \As {{A\kern-.5em / \kern.5em}}
\def \Ds {D\kern-.7em / \kern.5em}

\def \ks {k\kern-.5em /}
\def \ls {l\kern-.5em /}

\newcommand{\ci}[1]{}

\newcommand{\ba}{\begin{eqnarray}}
\newcommand{\ea}{\end{eqnarray}}
\newcommand{\bal}{\begin{align}}
\newcommand{\eal}{\end{align}}
\newcommand{\bay}[1]{\left(\begin{array}{#1}}
\newcommand{\eay}{\end{array}\right)}

\reversemarginpar
\newcommand{\hide}[1]{}

\makeatletter

\newlist{axioms}{enumerate}{2}
\setlist[axioms,1]{label=\textbf{A\arabic{axiomsi}.}, ref=A\arabic{axiomsi}}
\setlist[axioms,2]{label=\textbf{A\arabic{axiomsi}\rlap{\myEnumCounter{axiomsii}}.},%
                   ref=A\arabic{axiomsi}\myEnumCounter{axiomsii},%
                   align=parleft,%
                   leftmargin=0em,%
                   itemsep=1.4ex,%
                   before={\stepcounter{axiomsi}}}

\usepackage{tikz,pgf}
\usetikzlibrary{shapes}
\usetikzlibrary{calc}
\usetikzlibrary{decorations.pathmorphing}
\usetikzlibrary{decorations.pathreplacing,shapes.misc}
\usetikzlibrary{positioning}
\usetikzlibrary{arrows}
\usetikzlibrary{decorations.markings}

\tikzset{snake it/.style={decorate,decoration={snake,segment length=1.5mm, amplitude=.3mm}}}
\tikzset{biggerarrow/.style={
    decoration={markings,mark=at position 1 with {\arrow[scale=1.5]{>}}},
    postaction={decorate},
    shorten >=0.4pt}}
\tikzset{arrow at middle/.style={decoration={
    markings,
    mark=at position 0.5 with {\arrow{>}}}}}

\begin{document}

\begin{titlepage}

\begin{center}

\hfill
\vskip .2in

\textbf{\LARGE
Discussion of the Duality in Three Dimensional Quantum Field Theory
\vskip.3cm
}

\vskip .5in
{\large
Chen-Te Ma \footnote{e-mail address: yefgst@gmail.com} 
\\
\vskip 1mm
}
{\sl
Department of Physics and Center for Theoretical Sciences, \\
National Taiwan University,\\ 
Taipei 10617, Taiwan, R.O.C.
}\\
\vskip 1mm
\vspace{40pt}
\end{center}
\begin{abstract}
We discuss the duality in three dimensional quantum field theory at infrared limit. The starting point is to use a conjecture of a duality between the free fermion and the interacting scalar field theories at the Wilson-Fisher fixed point. The conjecture is useful for deriving various dualities in three dimensions to obtain a duality web. The study is also interesting for understanding the dualities, or equivalence of different theories from the perspective of the renormalization group flow. We first discuss the ``derivation'' without losing the holonomy. Furthermore, we also derive these dualities from the mean-field study, and consider the extension of the conjecture or dualities at finite temperature.

\end{abstract}

\end{titlepage}

\section{Introduction}
\label{1}
Duality is an equivalence of two theories from some non-trivial transformations. The duality is a useful tool to study theoretical physics because we could use different mathematical ways to see one physical phenomenon. Especially, we are particularly interested in dualities between the weakly coupled and strongly coupled field theories to study the strongly coupled field theories from the weakly coupled field theories. 

Three dimensional quantum field theory at infrared limit is particularly interesting because of time reversal symmetry anomaly in free Dirac fermion. The massless Dirac fermion at quantum level is not a well-defined theory and the massive Dirac fermion theory could be related to the Abelian Chern-Simons term so one way of finding a well-defined theory at infrared limit is to combine the massless Dirac fermion and the induced Abelian Chern-Simons term, which comes from the one-loop effective potential of the Dirac fermion \cite{Seiberg:2016rsg}. When one considers finite temperature, the Dirac fermion is not enough to be combined only the lowest non-trivial order term, the induced Abelian Chern-Simons term, to obtain the gauge symmetry in the effective theory. Thus, the resummation is necessary for restoring the gauge symmetry of the effective theory at the Wilson-Fisher fixed point. The theory at zero temperature is also conjectured to dual to the interacting scalar field theory at the Wilson-Fisher fixed point coupled to the $BF$ and Chern-Simons theories in the manifold with a chosen spin structure. The conjecture at zero temperature could exhibit the known particle-vortex duality \cite{Peskin:1977kp} so it should be a concrete evidence. An interesting fact is that the conjecture also gives a duality web to find the relation between the free theory and interacting theory \cite{Peskin:1977kp,Seiberg:2016gmd, Karch:2016aux}.

In this paper, we discuss the ``derivation'' (We use the ``derivation'' because one derives these dualities from a conjecture.) of the dualities in three dimensional quantum field theory. The ``derivation'' starts from one conjecture to derive other dualities. When one ``derives'' the dualities, one needs to integrate out gauge field to solve $da=db$, where $a$ and $b$ are gauge fields. The general solution of $da=db$ is $a=b+df$ with a globally defined $f$ if $a$ and $b$ satisfy the Dirac quantization condition. Roughly speaking, we could set $f=0$ or zero holonomy due to the gauge transformation. When a theory only has dynamical gauge fields, then the case is not problematic. If one theory has background gauge fields, then setting zero homonomy is not always doable because the gauge transformation could be performed only on the dynamical fields. However, we find that the non-trivial holonomy could be absorbed by the scalar field, and the non-trivial holonomy does not affect the dualities when a boundary term is absent, and all fields are globally defined. The mean-field study is a powerful tool to study the phase diagram so we consider the order parameter $\bar{\psi}\psi$ in the fermion side. The above studies only focus on the zero temperature. The phase diagram is particularly interesting when one considers the phase transition with respect to finite temperature. The difficulty of the conjecture at finite temperature is that it is necessary to include higher order terms beyond the induced Abelian Chern-Simons term. The inclusion of the higher order terms possibly affects the validity of the conjecture because of the kinetic term of a gauge field. Thus, only the induced Abelian Chern-Simons term with a temperature dependent coefficient is considered in our conjecture. Although the effective theory loses the gauge symmetry, the role of the temperature possibly be the same as the role of the Dirac fermion mass term at the Wilson-Fisher fixed point. The Dirac fermion mass term should vanish at the Wilson-Fisher fixed point, but we could do the perturbation to understand the dual operator, boson mass term. Thus, the duality at finite temperature and infrared limit still possibly works. 

We first ``derive'' dualities from the conjecture without ignoring the holonomy in Sec.~\ref{2}. The mean-field study of the duality is shown in Sec.~\ref{3}, and the dualities at finite temperature are exhibited in Sec.~\ref{4}. Finally, we give outlook and conclude in Sec.~\ref{5}. More details and related materials will be reported in a forthcoming paper \cite{HM}. 

\section{``Derivation'' of the Duality}
\label{2}
We first exhibit our notation to conveniently show the duality, and start from a conjecture of a duality between boson and fermion systems \cite{Seiberg:2016gmd} to derive the known duality for the particle-vortex duality of the boson fields at infrared limit without losing the holonomy.

\subsection{Notation}
Our notation for the action of the Abelian Chern-Simons with level one
\bea
S_{CS}[A]=\frac{1}{4\pi}\int d^3x\ \epsilon^{\mu\nu\rho}A_{\mu}\partial_{\nu}A_{\rho},
\eea
the action of the BF theory
\bea
S_{BF}[a ; A]=\frac{1}{2\pi}\int d^3x\ \epsilon^{\mu\nu\rho}a_{\mu}\partial_{\nu}A_{\rho},
\eea
the action of the scalar field theory
\bea
S_{scalar}[\phi; A]=\int d^3x\ \big(|(\partial_{\mu}+iA_{\mu})\phi|^2-\lambda|\phi|^4\big),
\eea
the action of the massless Dirac fermion theory
\bea
S_{fermion}[\psi; A]=\int d^3x\ i\bar{\psi}\gamma^{\mu}(\partial_{\mu}+iA_{\mu})\psi,
\eea
the partition function of the massless Dirac fermion
\bea
Z_{fermion}[A]=\int D\psi\ \exp\big(iS_{fermion}[\psi ;A]\big),
\eea
the partition function of the scalar field theory
\bea
Z_{scalar}[A]=\int D\phi\ \exp\big(iS_{scalar}[\phi ;A]\big),
\eea
in which we used the metric $\eta_{\mu\nu}=\mbox{diag}(1, -1, -1,\cdots, -1)$ to do contraction.

\subsection{Conjecture}
We start from a conjecture:
\bea
Z_{fermion}[A]\exp\bigg(-\frac{i}{2}S_{CS}[A]\bigg)=\int D\phi Da\ \exp\big(i S_{scalar}[\phi; a]+iS_{CS}[a]+iS_{BF}[a; A]\big)
\nn\\
\eea
in three dimensions. The massless Dirac fermion is not gauge invariant at quantum level so it is necessary to include the induced Abelian Chern-Simons theory with level one to restore gauge symmetry.

The electric current of the fermion $i\bar{\psi}\gamma^0\psi$ could be related to the flux current of $a$, $(1/2\pi)\epsilon^{0\nu\rho}\partial_{\nu}a_{\rho}$. This leads that the monopole of $a$ could give $A$ charge one for the fermion. The Chern-Simons term for $a$ could generate $a$ charge one, and the $a$ charge could be neutralized by adding a complex boson $\phi$ because the angular momentum of $\phi$ is one half. Thus, the state has the same angular momentum and the dynamical gauge field also has the same $A$ charge as in the fermion field. The conjecture could be correct because we consider $e^2\rightarrow\infty$ or the infrared limit, then the kinetic term of the gauge field could be ignored.

We could rewrite the conjecture:
\bea
\int D\psi DA\ \exp\bigg(iS_{fermion}[\psi ; A]-\frac{i}{2}S_{CS}[A]-iS_{BF}[A ;C]\bigg)=e^{iS_{CS}[C]}\int Df\ Z_{scalar}[C+df],
\nn\\
\eea
in which we equivalently use $A=C+df$ when we do path integration. We could use the action of the time reversal to obtain
\bea
\int D\psi DA\ \exp\bigg(iS_{fermion}[\psi ; A]+\frac{i}{2}S_{CS}[A]+iS_{BF}[A ;C]\bigg)=e^{-iS_{CS}[C]}\int Df\ Z_{scalar}[C+df].
\nn\\
\eea

\subsection{Particle-Vortex Duality of Boson Fields}
We derive a particle-vortex duality \cite{Peskin:1977kp} of boson fields from the conjecture, and start from
\bea
e^{-iS_{CS}[C]}\int D\psi DA\ \exp\bigg(iS_{fermion}[\psi ; A]-\frac{i}{2}S_{CS}[A]-iS_{BF}[A ;C]\bigg)=\int Df\ Z_{scalar}[C+df].
\nn\\
\eea
We could add the $BF$ term to get
\bea
&&\int Da\ \exp\bigg(-iS_{CS}[a]+iS_{BF}[a ; A]\bigg)\int D\psi D\tilde{a}\ \exp\bigg(iS_{fermion}[\psi ; \tilde{a}]-\frac{i}{2}S_{CS}[\tilde{a}]-iS_{BF}[\tilde{a} ;a]\bigg)
\nn\\
&=&\int Da\ \exp\bigg(iS_{BF}[a ; A]\bigg)\int Df\ Z_{scalar}[a+df],
\eea
then we could integrate out the dynamical gauge field $a$, which is equivalent to getting $a=A-\tilde{a}+dg$ to obtain
\bea
&&\int D\psi D\tilde{a}\ \exp\bigg(iS_{fermion}[\psi ; \tilde{a}]
+\frac{i}{2}S_{CS}[\tilde{a}]-iS_{BF}[\tilde{a} ;A]+iS_{CS}[A]\bigg)
\nn\\&=&\int Da\ \exp\bigg(iS_{BF}[a ; A]\bigg)\int Df\ Z_{scalar}[a+df].
\eea
Therefore, we obtain
\bea
\int Df\ Z_{scalar}[-A+df]=\int Da\ \exp\bigg(iS_{BF}[a ; A]\bigg)\int Df\ Z_{scalar}[a+df].
\eea
Since the gauge transformation of $a$ is $d\lambda$, the shift of $a$, $a\rightarrow a-df$ does not modify the $BF$ term, we could use $\phi\rightarrow\phi\cdot\exp(-if)$ to absorb the $df$ for the bosonic parts. Thus, we could show that the particle-vortex duality of the bosonic fields is not affected by the holonomy:
\bea
Z_{scalar}[-A]\sim\int Da\ \exp\bigg(iS_{BF}[a ; A]\bigg)Z_{scalar}[a].
\eea
Other dualities \cite{Seiberg:2016gmd} are also not affected by the holonomy from the same way.

\section{The Mean-Field Study of the Duality}
\label{3}
We use the order parameter $\bar{\psi}\psi$ in the fermion side to study the duality \cite{Karch:2016aux} at infrared limit. The four fermion interacting term at infrared limit or $g^2\rightarrow 0$ could be considered to recover the massless Dirac fermion:
\bea
S_{gn}=\int d^3x\ \bigg(\bar{\psi}\gamma^{\mu}\big(i\partial_{\mu}-A_{\mu}\big)\psi+\frac{g^2}{2}\big(\bar{\psi}\psi\big)^2\bigg).
\eea
We introduce an auxiliary field $m$ to rewrite the action
\bea
\int d^3x\ \bigg(\bar{\psi}\gamma^{\mu}\big(i\partial_{\mu}-A_{\mu}\big)\psi-\frac{1}{2g^2}m^2-m\bar{\psi}\psi\bigg).
\eea
When taking the infrared limit, the auxiliary field $m$ approaches a constant or fixed value from the mean-field analysis. Hence, the fermion theory in the conjecture is written as
\bea
Z_{fermion}[A]\cdot e^{-\frac{i}{2}S_{CS}[A]}\delta\big(\bar{\psi}\psi-\hat{m}\big),
\eea
where $\hat{m}$ is a constant.

To derive the dual scalar field theory, we rewrite the conjecture
\bea
&&\lim_{g\rightarrow 0,\alpha\rightarrow\infty}\int D\mu D\sigma\ \Bigg\{\Bigg\lbrack\ \int D\psi\ \exp\bigg\lbrack i\int d^3x\ \bigg(\bar{\psi}\gamma^{\mu}\big(i\partial_{\mu}-A_{\mu}\big)\psi-\frac{1}{2g^2}\sigma^2-\sigma\bar{\psi}\psi\bigg)\bigg\rbrack
\nn\\
&&\times \exp\bigg(-\frac{i}{2}S_{CS}[A]+i\mu( \sigma-\tilde{m})\bigg)\Bigg\rbrack
\nn\\
&-&\Bigg\lbrack\int D\phi Da\ \exp\bigg\lbrack i\int d^3x\ \bigg(|(\partial_{\mu}+ia_{\mu})\phi|^2-\sigma|\phi|^2+\frac{1}{4\alpha}\sigma^2\bigg)\bigg\rbrack
\nn\\
&&\times\exp\bigg(iS_{CS}[a]+iS_{BF}[a; A]+i\mu(\sigma-\tilde{m})\bigg)\Bigg\rbrack\Bigg\}\approx 0,
\nn\\
\eea
where $\tilde{m}$ is determined by the mean-field analysis. The conjecture could go back to the original conjecture if we integrate out $\tilde{m}$. Now we only use the mean-field for the effects of $\tilde{m}$.

We begin to integrate out $\mu$, then we could obtain
\bea
&&\lim_{g\rightarrow 0,\alpha\rightarrow\infty}\ \Bigg\lbrack\ \int D\psi\ \exp\bigg\lbrack i\int d^3x\ \bigg(\bar{\psi}\gamma^{\mu}\big(i\partial_{\mu}-A_{\mu}\big)\psi-\frac{1}{2g^2}\tilde{m}^2-\tilde{m}\bar{\psi}\psi\bigg)\bigg\rbrack
\cdot \exp\bigg(-\frac{i}{2}S_{CS}[A]\bigg)\Bigg\rbrack
\nn\\
&\approx&\Bigg\lbrack\int D\phi Da\ \exp\bigg\lbrack i\int d^3x\ \bigg(|(\partial_{\mu}+ia_{\mu})\phi|^2-\tilde{m}|\phi|^2+\frac{1}{4\alpha}\tilde{m}^2\bigg)\bigg\rbrack\cdot
\exp\bigg(iS_{CS}[a]+iS_{BF}[a; A]\bigg)\Bigg\rbrack.
\nn\\
\eea
Then we use the mean-field analysis of the fermion fields to determine $\tilde{m}$. Thus, we could have
\bea
&&\int D\psi\ \exp\bigg\lbrack i\int d^3x\ \bigg(\bar{\psi}\gamma^{\mu}\big(i\partial_{\mu}-A_{\mu}\big)\psi\bigg)\bigg\rbrack
\cdot \exp\bigg(-\frac{i}{2}S_{CS}[A]\bigg)\delta\big(\bar{\psi}\psi-\hat{m}\big)
\nn\\
&\approx&\int D\phi Da\ \exp\bigg\lbrack i\int d^3x\ \bigg(|(\partial_{\mu}+ia_{\mu})\phi|^2-\tilde{m}|\phi|^2\bigg)\bigg\rbrack\cdot
\exp\bigg(iS_{CS}[a]+iS_{BF}[a; A]\bigg),
\nn\\
\eea
where $\hat{m}=-\tilde{m}/g^2$ is determined form the mean-field analysis at the limit $g\rightarrow 0$.

Our conclusion is that the mean-field analysis of the fermion theory could dual to the massive scalar field theory without the interacting term (due to mean-field) at infrared limit. The kinetic term of the gauge field also does not appear to affect the validity of the conjecture because we consider the Wilson-Fisher fixed point.

\section{Duality at the Finite Temperature}
\label{4}
The free fermion at finite temperature and infrared limit could be related to the induced Abelian Chern-Simons theory with a temperature dependent coefficient. The temperature dependent coefficient breaks the gauge symmetry of the effective theory. One way of restoring the gauge symmetry in the effective theory is to do resummation to all orders with respect to number of the gauge field. The inclusion of the higher order terms in the effective theory may break the dualities at finite temperature so we only keep the leading order term. At the first, if we only keep the induced Abelian Chern-Simons term, the form of the dualities is the same as the dualities at zero temperature so we could have the same physical conclusion as in the case of the zero temperature. If we turn on the positive fermion mass, the theory at infrared limit is still trivial, and the case corresponds to the positive boson mass. The negative fermion mass case gives the Abelian Chern-Simons theory with a temperature dependent coefficient, and the negative boson mass case also gives the Abelian Chern-Simons theory with a temperature dependent coefficient. The correspondence is not only valid at the low temperature. At high temperature limit, we find that the induced Abelian Chern-Simons term at finite temperature vanishes, and the symmetry is also restored for the negative boson mass in the scalar field theory to give the trivial theory for the boson part. Thus, we do not find the obvious violation for the conjecture at finite temperature.

\section{Conclusion and Outlook}
\label{5}
We discuss the duality in three dimensional quantum field theory at infrared limit. Our starting point is to understand the ``derivation'' of the duality with the holonomy clearly. Thus, we ``derive'' the particle-vortex duality \cite{Peskin:1977kp} for the bosons without losing the holonomy. We find that the result is not modified by the holonomy. The result could be extended to other dualities \cite{Seiberg:2016gmd}.

We are also interested in studying the duality from the point of view of the phase diagram so we discuss the mean-field analysis and finite temperature in the duality. To consider the mean-field analysis, we introduce the four fermion interacting term, which could go back to the free fermion theory at infrared limit. The dual boson theory is a massive scalar field after we consider the order parameter $\bar{\psi}\psi$ or the mean-field analysis in the fermion side. The difficulty of introducing temperature is that the free fermion theory at infrared limit needs to do resummation to all orders to have gauge symmetry in the effective theory. Our consideration of the duality is only guaranteed at infrared limit so we only keep the induced Abelian Chern-Simons term with a temperature dependent coefficient. 

The effective theory at finite temperature loses the gauge symmetry because of truncation. If we start from the non-truncated theory, the issue of the gauge symmetry does not appear although the theories may not have dualities to connect each other. To get the dualities at finite temperature, we could use perturbation to examine. Thus, the dualities at finite temperature possibly works even if the effective theory loses the gauge symmetry.

\section*{Acknowledgments}
The author would like to thank Chang-Tse Hsieh for his useful discussion. Especially, the author would like to thank Nan-Peng Ma for his suggestion and encouragement. 

\end{document}